\documentclass[preprint,preprintnumbers,amsmath,amssymb]{revtex4}

\usepackage{graphicx}
\usepackage{dcolumn}
\usepackage{bm}
\usepackage{color}

\begin{document}

\preprint{APS/123-QED}

\title{Novel quantum phase of the chromium spinel oxide HgCr$_{\rm 2}$O$_{\rm 4}$\\ in high magnetic fields}

\author{Shojiro Kimura$^{\rm 1}$, Shusaku Imajo$^{\rm 2}$, Masaki Gen$^{\rm 2}$, Tsutomu Momoi$^{\rm 3}$, Masayuki Hagiwara$^{\rm 4}$, Hiroaki Ueda$^{\rm 5}$, Yoshimitsu Kohama$^{\rm 2}$}
\affiliation{$^{\rm 1}$Institute for Materials Research, Tohoku University, Katahira 2-1-1, Sendai 980-8577, Japan\\
$^{\rm 2}$Institute for Solid State Physics, University of Tokyo, Kashiwa 277-8531, Japan\\
$^{\rm 3}$Condense Matter Theory Laboratory, RIKEN, Wako, Saitama 351-0198, Japan\\
$^{\rm 4}$Center for Advanced High Magnetic Field Science, Graduate School of Science, Osaka University, Toyonaka Osaka 560-0043, Japan\\
$^{\rm 5}$Department of Chemistry, Graduate School of Science, Kyoto University, Kitashirakawa Oiwake-cho, Sakyo-ku, Kyoto 606-8502, Japan}

\date{\today}

\begin{abstract}
In this study, we have performed the magnetocaloric effect and the specific heat measurements of chromium spinel oxide HgCr$_2$O$_4$, wherein the magnetic Cr$^{3+}$ ions form a highly frustrated pyrochlore lattice with significant spin-lattice coupling. In addition to the known magnetic-field-induced phases, our thermodynamic measurements detect a novel quantum phase just before the saturation of the magnetization, which has not been expected from the classical theories of the pyrochlore lattice antiferromagnet with spin-lattice coupling. Based on recent theoretical model calculation, we discuss the possibility of a  spin nematic state appearing for this quantum phase.  

\end{abstract}

\maketitle
The spin nematic state, which has long been discussed for magnets with frustrated interactions, is a spin analog of the nematic state in a liquid crystal\cite{Blume,Andreev,Chubukov,Shannon,Lacroix}. In this unconventional magnetic state, the fluctuations of spins, behaving like rod- or disk-shaped molecules, spontaneously select a preferred direction without conventional magnetic order\cite{Lacroix,Lauchli,Smerald}.  
Under an external magnetic field, the spin nematic state is defined by finite values of transverse spin quadrupolar order parameters ${\langle}S_{i}^{-}S_{j}^{-}+h.c.{\rangle}$ with no transverse dipolar order, i.e. ${\langle}S_{i}^{x}{\rangle}={\langle}S_{i}^{y}{\rangle}= 0$\cite{Shannon,Orlova}. A promising route to obtain the spin nematic state is  the condensation of magnon bound pairs, which leads to the finite quadrupolar order parameters\cite{Shannon,Momoi}. As a realistic system, which is able to stabilize the two-magnon bound state, a square lattice  and one-dimensional chain magnets with competing ferromagnetic and antiferromagnetic exchange interactions was first suggested\cite{Chubukov,Shannon,Hikihara,Zhitomirsky,Ueda}. The competition between the exchange interactions suppresses the hopping of individual magnons, while the ferromagnetic interaction, which acts as an attractive force between magnons, stabilizes the bound states of multiple magnons, causing the spin nematic ground state. Based on this prediction, the searches for the spin nematic state have been intensively conducted in several frustrated magnets with spin $S$ = 1/2, such as LiCuVO$_4${\cite{Svistov,Orlova,Mourigal, Gen}, Cu$_3$V$_2$O$_7$(OH)$_2$$\cdot$2H$_2$O (volborthite)\cite{Yoshida,Kohama}, and BaCdVO(PO$_4$)$_2$\cite{Skoulatos}.  
 Moreover, a new method to realize the spin nematic order by condensation of two-magnon bound states was recently proposed for an antiferromagnet on the highly frustrated pyrochlore lattice with spin-lattice coupling, which is fulfilled in the chromium spinel oxides $A$Cr$_2$O$_4$ ($A=$ Mg, Zn, Cd, and Hg)\cite{Takata}. It was demonstrated theoretically that the effective biquadratic spin interaction, which has been known to stabilize the spin nematic state\cite{Lacroix,Lauchli,Smerald,Tsunetsugu}, is derived from the spin-lattice coupling\cite{penc1,penc2,Motome}. Takata $\it{et. al.}$ demonstrated that this biquadratic interaction provides a significant energy gain for the two-magnon bound state on the pyrochlore lattice to lead the spin nematic ground state just below the saturation field\cite{Takata}.

In $A$Cr$_2$O$_4$, which belongs to the cubic $Fd\bar{3}m$ space group, magnetic Cr$^{3+}$ ions with $S$ = 3/2 form the pyrochlore lattice, which is composed of a three-dimensional arrangement of corner-sharing tetrahedra. Because the strong spin frustration from the lattice geometry causes vast degeneracy in the ground state, both classical $S=\infty$ and quantum $S=1/2$  pyrochlore antiferromagnets with the nearest neighbor Heisenberg interaction are believed to show no long range order at any temperature\cite{Moessner,Canals}. However, in $A$Cr$_2$O$_4$, the spin ordering is allowed to occur due to the weak further neighbor exchange interaction and the spin-lattice coupling, causing lattice distortion at the ordering temperature to relieve the ground state degeneracy\cite{Kino,Plumier, Rovers,Yamashita,Tchernyshyov1,Tchernyshyov2}. Under the external magnetic fields $H$, $A$Cr$_2$O$_4$ shows successive field-induced phase transitions, including a distinctive half-magnetization plateau phase with three up and one down spin configuration in a tetrahedron\cite{ueda1,ueda2,Mitamura,Kojima,miyata1,miyata2,miyata3,miyata4,miyata5}. The magnetization curve in HgCr$_2$O$_4$, of which the exchange interaction is the weakest among $A$Cr$_2$O$_4$, is shown in Fig. 1\cite{ueda2,Kimura1,Kimura2}. The magnetization $M$, which gradually increases in the low field region, jumps to the plateau with half of the saturation magnetization, at $H_{c1}$ = 10 T. Then, $M$ starts to increase again above $H_{c2}$ = 27 T and shows a kink at $H_{c3}$ = 36 T with a hysteresis, refrecting the first order nature of the transition. This magnetization process is accompanied by deformations of the crystal lattice, which cause large changes in the nearest neighbor exchange interactions. Via the transition to the plateau phase above $H_{c1}$, the crystal symmetry of HgCr$_2$O$_4$ changes to the distorted cubic $P4_332$\cite{Matsuda}, causing an imbalance in the exchange interactions in the tetrahedron,  which stabilizes the magnetization plateau\cite{Kimura1,Kimura2}. Above $H_{c2}$, where the canted 3:1 magnetic structure results in the gradual increase of $M$, this imbalance is gradually diminished as the magnetic field $H$ is increased. It then disappears at $H_{c3}$, indicating restoration of the high symmetry $Fd\bar{3}m$ lattice with no distortion\cite{Kimura1,Kimura2}. The interplay between the lattice and the magnetism in $A$Cr$_2$O$_4$ is accounted for the effective spin Hamiltonian, derived by integrating out the lattice degrees of freedom\cite{penc1,penc2,Motome}. Its most simple form is the one with the bilinear and the aforementioned biquadratic interactions with an constant $b$: 
\begin{eqnarray}
{\mathcal H} = J \sum_{i,j} \left\{ \bm{S}_{i}\bm{S}_{j}-b(\bm{S}_{i}\bm{S}_{j})^{2} \right\}-g\mu_B\bm{H}\sum_{i}\bm{S}_{i}.
\end{eqnarray}
Our previous work demonstrated that classical mean field calculations, based on the magnetoelastic Hamiltonian with several spin-lattice coupling constants for each normal tetrahedral distortion mode, reproduce the overall field dependencies of $M$ and the exchange interactions, deduced from the ESR measurements\cite{Kimura2}. However, there is an unresolved discrepancy between the experiments and the mean field calculations. While the classical calculation indicates the saturation of the $M$ at $H_{c3}$, the experimentally observed $M$ still gradually increases above $H_{c3}$. It is noteworthy that there is a tiny peak in the $dM/dH$ curve at $H_{c4}$ = 39 T as shown in Fig. 1. This indicates the existence of a magnetic phase, which is not anticipated from the classical model, for $H_{c3}<H<H_{c4}$. In this unanticipated phase, the crystal lattice has no distortion while the magnetization does not saturate but still increases as the magnetic field is increased\cite{Kimura1}.  We consider that this phase is an unconventional state, because, even in the magnetic field, a pyrochlore antiferromagnet with no lattice distortion possesses vast degeneracy in the ground state\cite{penc1}, which is expected to strongly disturb conventional magnetic order. As a reasonable candidate for this phase, the spin nematic state has been recently proposed\cite{Takata}. To search for the possible spin nematic phase, we have performed thermodynamic experiments in high magnetic fields. The thermodynamic methods can detect phase transitions characterized by any type of order parameter, including the long range order of the spin quadrupole moment i.e., spin nematic order.
 In this study, we have measured the high-field specific heat and magnetocaloric effect of HgCr$_{\rm 2}$O$_{\rm 4}$ in pulsed magnetic fields of up to 42 T to gain thermodynamic evidence for the existence of ordering in the field region between $H_{c3}$ and $H_{c4}$.

Polycrystalline samples of HgCr$_2$O$_4$ were prepared by thermal decomposition of Hg$_2$CrO$_4$\cite{ueda2}. We have measured the temperature $T$ dependence of the specific heat  $C_p$ in HgCr$_2$O$_4$ at the high magnetic fields of 15.8, 26.9, 34.8, and 38.5 T using the quasi-adiabatic method under a flat-top pulsed field\cite{Imajo,Kohama1}.  The $C_p$ at 0 T and 8.0 T was measured using the conventional thermal relaxation method by utilizing a commercial measurement system (Quantum Design, PPMS). To measure the field dependence of specific heat under the quasi-isothermal condition, the ac-calorimetry of up to 40 T was performed at 250 Hz\cite{Kohama2}. The magnetocaloric effect (MCE) measurements were also performed with high magnetic fields of up to 42 T using the same setup without applying AC heat. By adjusting the amount of helium gas in the sample space, the MCE data were obtained in different thermal environments: quasi-adiabatic or nearly isothermal conditions.

Figure 2(a) shows the MCE curve $T(H)$ started at 0.8 K under quasi-adiabatic condition in the up-sweep of the field pulse. The $T$($H$) exhibits a sharp dip at $H_{c1}$, and shows a maximum  around 17 T in the plateau phase, which roughly corresponds to the center of the magnetization plateau.Then, the $T$($H$) decreases as the magnetic field is increased and exhibits a minimum  
at $H_{c2}$. Above $H_{c2}$, a kink of $T$($H$) is observed at $H_{c3}$. Between $H_{c2}$ and $H_{c3}$, a small dip of $T$($H$) is found at arond $H_{c2'}$ = 31 T. An anomaly around 31 T in the ${dM/dH}$ curve, which has been overlooked in previous papers\cite{ueda2,Kimura1,Kimura2}, is also shown in Fig. 1. This anomaly suggests some weak but sudden change in the spin structure. However its detail is not clear at present.
In ref. 47, the positive MCE by entering the plateau phase has been predicted. According to the thermodynamic equations, the field derivative of the $T$($H$), observed under an adiabatic condition, is given as\cite{Rossi,Schmidt}:
\begin{eqnarray}
\biggl(\frac {\partial T}{\partial H} \biggr)_S = -\frac {T}{C_p}\biggl(\frac {\partial M}{\partial T} \biggr)_H. 
\end{eqnarray}
Thus, the sign of ($\partial T$/$\partial H$)$_S$ is determined by that of ($\partial M$/$\partial T$)$_H$ because both $C_p$ and $T$ are positive. The sharp dip at $H_{c1}$ and the subsequent increase in $T$($H$), observed in our experiments, are consistent with the ($\partial M$/$\partial T$)$_H$ obtained from the classical Monte Carlo simulations and analytical calculations based on the magnetoelastic Hamiltonian with the weak ferromagnetic third nearest neighbor interaction $J_3$ in the supplementary materials for ref. 47. However, while the calculated ($\partial M$/$\partial T$)$_H$ suggests a monotonic increase in $T$($H$) in the entire plateau region, the experimental $T$($H$) starts to decrease of above 17 T. The calculation indicates that the change in the sign of the ($\partial M$/$\partial T$)$_H$ from negative to positive occurs at $H_{c2}$, namely, just at the end of the plateau phase. In contrast, the temperature dependence of the experimental $M$, reported in ref. 33, displayed the sign change of the ($\partial M$/$\partial T$)$_H$ within the plateau phase, therefore agreeing with our MCE result. The classical Monte Carlo simulation with $J_3$ = 0 by Shannon $et$ $at$.\cite{Shannon2} also suggested that the sign change occurs at the magnetic field below $H_{c2}$. We suspect that the discrepancy between the calculation in the supplementary materials for ref. 47 and the experimental results may likely comes from the assumed ferromagnetic $J_3$, which is incompatible with the magnetic structure of the $P4_332$ symmetry in the plateau phase, and a possible quantum effect. A more apparent discrepancy between the calculations and the experimental result is observed around $H_{c3}$. The calculation suggested a deep and sharp minimum of $T$($H$) at $H_{c3}$ and also an increase of $T$($H$) above $H_{c3}$, whereas the experimental $T$($H$) is almost constant above $H_{c3}$. 
The strong anomaly, derived from the classical calculation, is associated to the energy-gap-closing of the localized spin wave mode\cite{Rossi}, which is characteristic of pyrochlore lattice magnets, at the saturation field and the subsequent gap-reopening.  The gap-closing at $H_{c3}$ was detected from the previous high-field ESR measurements in HgCr$_2$O$_4$\cite{Kimura1}. However, when the spin nematic state is realized, the system still remains to be gapless above $H_{c3}$ because of an emergence of the Goldstone mode for the quadrupolar order parameter by breaking of rotational symmetry inherent in the spin nematic order\cite{Lauchli,Smerald}.
This gapless nature of the spin nematic state may suppress the strong anomaly of $T$($H$) at $H_{c3}$ as experimentally observed.

To clarify the temperature dependence of the magnetic-field-induced transitions, we performed the MCE measurements at several temperatures in a nearly isothermal condition (Fig. 2(b)). Note that, these MCE curves were taken under the nearly isothermal condition with a small amount of $^3$He gas. The anomalies observed in this condition are more smeared out than those in quasi-adiabatic condition due to the heat leak via the helium gas around the sample. When the amount of gas is large, the measurement conditions are closer to the isothermal limit, and the transitions become less visible, as shown in the MCE data obtained at 1.7 K in Fig. 2(b). As shown in the inset in Fig. 2(a), the MCE result under a condition intermediate between adiabatic and nearly-isothermal at around 2 K clearly exhibits sharp dips both at $H_{c3}$ $\simeq$ 37 T and $H_{c4}$ $\simeq$ 40 T.This results unambiguously indicates an existence of the new high field phase between $H_{c3}$ and $H_{c4}$. Together with the MCE data obtained in the quasi-adiabatic and intermediate conditions and the $C_p(T,H)$ data shown below, the MCE data in the nearly isothermal condition were used to construct the $(B,T)$ phase diagram in Fig. 4.

Figure 3(a) shows the temperature dependencies of the specific heat $C_p$. At zero magnetic field, a peak of the $C_p$, which is indicative of a magnetic long-range order, is observed around the reported N$\rm{\acute{e}}$el temperature of 5.8 K. This peak slightly shifts to lower temperature in the magnetic field at 8 T. In the plateau phase at 15.8 T, a sharp peak of $C_p$ is observed at a higher temperature. The peak broadens above $H_{c2}$, and then the broad peak is still observed at 38.5 T, which is well above $H_{c3}$. It is important to note that the peak in $C_p$ could not fit to the Schottky-type heat anomaly related to a non-cooperative thermal excitation in a gapped system. This suggests that the peak is attributed to the critical behavior in a phase transition.
Figure 3(b) shows the field dependence of $C_{\rm {p}}$ observed at 1.3 K. The anomalies are observed at $H_{c1}$, $H_{c2}$ and $H_{c3}$. In contrast to the fact that the specific heat of an antiferromagnet often shows a pronounced peak at a saturation field because of the subsequent reopening of the spin wave gap as reported in ref. \cite{Kohama3}, $C_{\rm {p}}$ in HgCr$_2$O$_4$ exhibits only the tiny anomaly at $H_{c3}$ and the gradual increase above $H_{c3}$, implying the existence of the subsequent phase transition at the higher magnetic field. These behaviors are also in contrast to the sharp anomaly in $dM/dH$ at $H_{c3}$. The observed $C_{\rm {p}}$ suggests that the $H_{c3}$ is not the saturation field, and a gapless ordered phase exists between $H_{c3}$ and $H_{c4}$.
The field-temperature phase diagram of HgCr$_2$O$_4$, obtained from this study, is shown in Fig. 4. As deduced from previous magnetization measurements in pulsed magnetic fields \cite{ueda2}, the plateau phase is expanded to a higher temperature region than the antiferromagnetic phase below $H_{c1}$ because of the stabilization by thermal fluctuation. The phase boundary between the paramagnetic and the plateau phases is suggested to be located at higher temperatures compared to the temperature reported from the magnetization measurements. This is possibly because the magnetization measurement does not consider the magnetocaloric effect but assumes isothermal conditions. 

As mentioned above, the spin nematic state has been theoretically proposed for the magnetic phase just below $H_{c4}$\cite{Takata}. This spin nematic state is stabilized in the pyrochlore lattice antiferromagnet owing to the interplay of the spin-lattice coupling, quantum fluctuation, and geometrical frustration. In the Heisenberg-type pyrochlore lattice magnet, the lowest-energy magnon states in a fully spin-polarized region above the saturation field are localized with flat dispersions owing to the geometrical configuration of the lattice\cite{Zhitomirsky,Zhitomirsky2}. Therefore, the hopping of the single lowest-energy magnon is prohibited, thereby preventing kinetic energy gain. Meanwhile the spin-lattice coupling in $A$Cr$_2$O$_4$ brings the effective biquadratic interaction in Eq. (1). This biquadratic interaction contains a term $(S_{i}^{-})^2 (S_{j}^{-})^2+h.c.$, which induces hopping of a two-spin flipped state on a Cr$^{3+}$ site to the nearest neighbor sites. Because of this hopping term, the two-magnon bound state becomes stable and has a lower energy than two independent localized magnons in the fully polarized spin state\cite{Takata}. Decreasing the magnetic field in the fully polarized spin state causes softening of the two-magnon bound state. Then, the bound state touches to the ground state at the saturation field, resulting in its condensation, which leads to the spin nematic state. Because the dispersion curve of the two-magnon bound state has a minimum at $\bm k =$ (0 0 0) in the reciprocal lattice space, where $\bm k$ is the center-of-mass wave vector, the resulting spin nematic state is characterized by the ferro-quadrupolar order. Thus, the spin nematic order causes no lattice distortion via the spin-lattice coupling. This is consistent with the absence of the symmetry lowering in the lattice from the $Fd\bar{3}m$ symmetry above $H_{c3}$ in HgCr$_2$O$_4$, as suggested from the previous ESR experiments\cite{Kimura1}. From the theoretical calculation, Takata $et$ $al.$ found that the spin nematic order can appear with relatively weak $b$ $>$ 0.043\cite{Takata}. This is also consistent with the reported value of $b_A$ = 0.05 in HgCr$_2$O$_4$\cite{Kimura2}, which is a biquadratic interaction constant for the fully symmetric $A_1$ mode. According to the calculation, the spin-component distribution, of which the projection to the plane perpendicular to the magnetic field has an ellipsoid shape, is aligned owing to the spin nematic order, thereby breaking the rotational symmetry around field\cite{Takata}. From our experimental results that suggest the existence of the gapless magnetic ordered phase between $H_{c3}$ and $H_{c4}$, we consider that such a spin nematic order most probably appears in HgCr$_2$O$_4$. The existence of an unknown magnetic phase between the 3:1 canted and the fully polarized spin states was also reported for other chromium spinel oxides, CdCr$_2$O$_4$ and ZnCr$_2$O$_4$, from high-field experiments\cite{miyata2,miyata3,miyata4,Nakamura,Zherlitsyn,Kimura3}. It is possible that the appearance of the spin nematic phase is of universal nature in $A$Cr$_2$O$_4$. To gain more detailed microscopic information of the phase between $H_{c3}$ and $H_{c4}$, nuclear magnetic resonance measurements in high magnetic fields are desirable. 	

In summary, we have performed the high-field thermodynamic measurements of the chromium spinel oxide HgCr$_2$O$_4$ in pulsed magnetic fields of up to 40 T. From the temperature and field dependences of the specific heat, we observed the appearance of the new magnetic phase between the 3:1 canted and the fully polarized spin states. We consider that the spin nematic state, which is stabilized due to the interplay of spin-lattice coupling, quantum fluctuation, and geometrical frustration in the pyrochlore lattice antiferromagnet, is realized in this magnetic phase.

The experimental work was performed under the Visiting Researcher's Program of the Institute for Solid State Physics, The University of Tokyo. This work was in part supported by the Grant-in-Aid for Scientific Research (Grant No. 121H01026 and No. 22H00101) from MEXT Japan.

\clearpage

\begin{figure}
\includegraphics[width=13cm,clip]{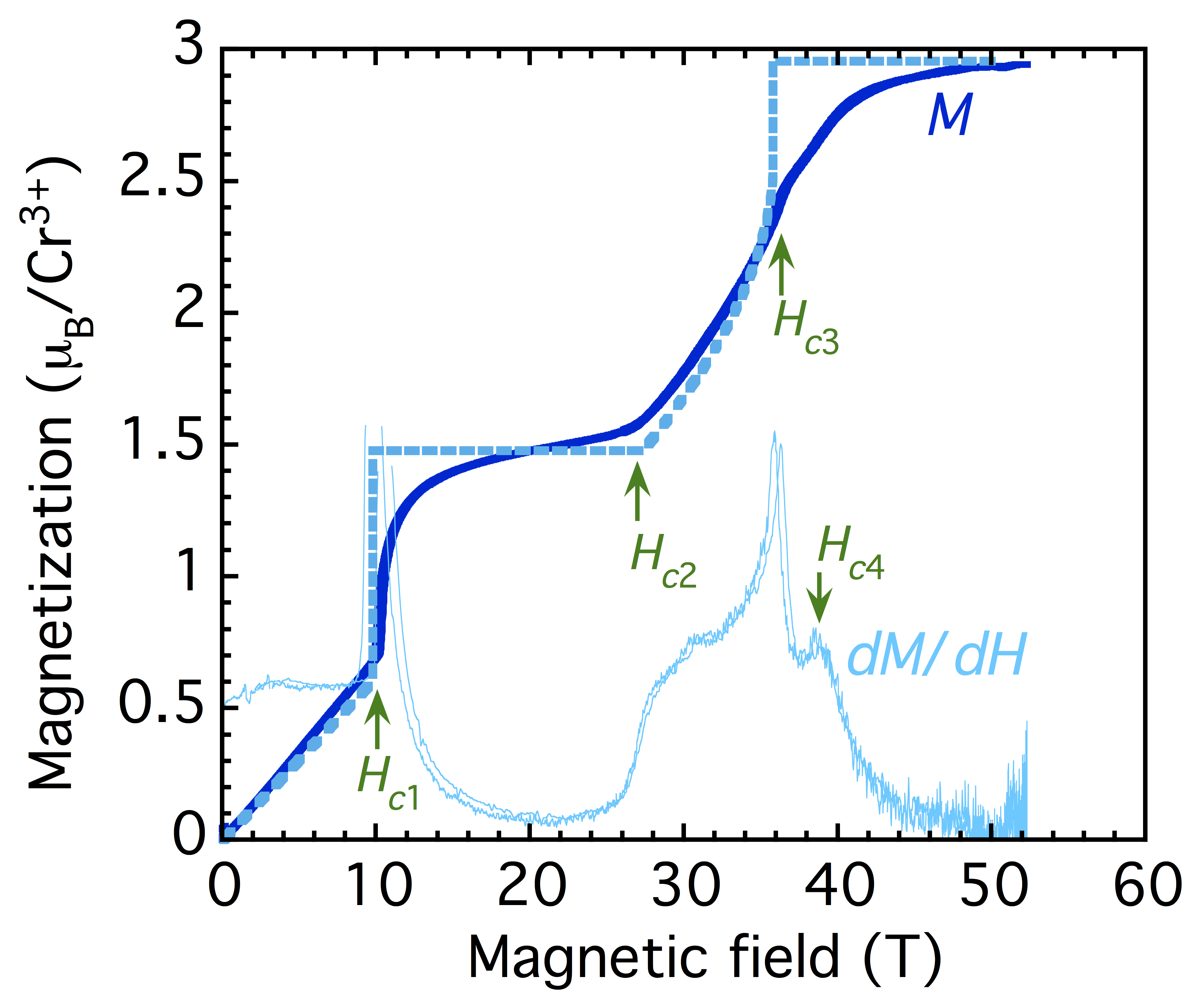}
\caption{High field magnetization curve of a polycrystalline sample of HgCr$_2$O$_4$, observed at 1.3 K. Solid bold curves are the experimental magnetization ($M$) curve, and a dashed curve is the theoretical curve obtained from the classical mean field calculation based on the magnetoelestic Hamiltonian. Thin solid curves are the differential magnetization ( $dM$/$dH$) curve. The experimental and theoretical data are from \cite{Kimura1} and \cite{Kimura2}, respectively.}
\label{Frequency-field plot}
\end{figure}

\pagebreak
\begin{figure}
\includegraphics[width=15cm,clip]{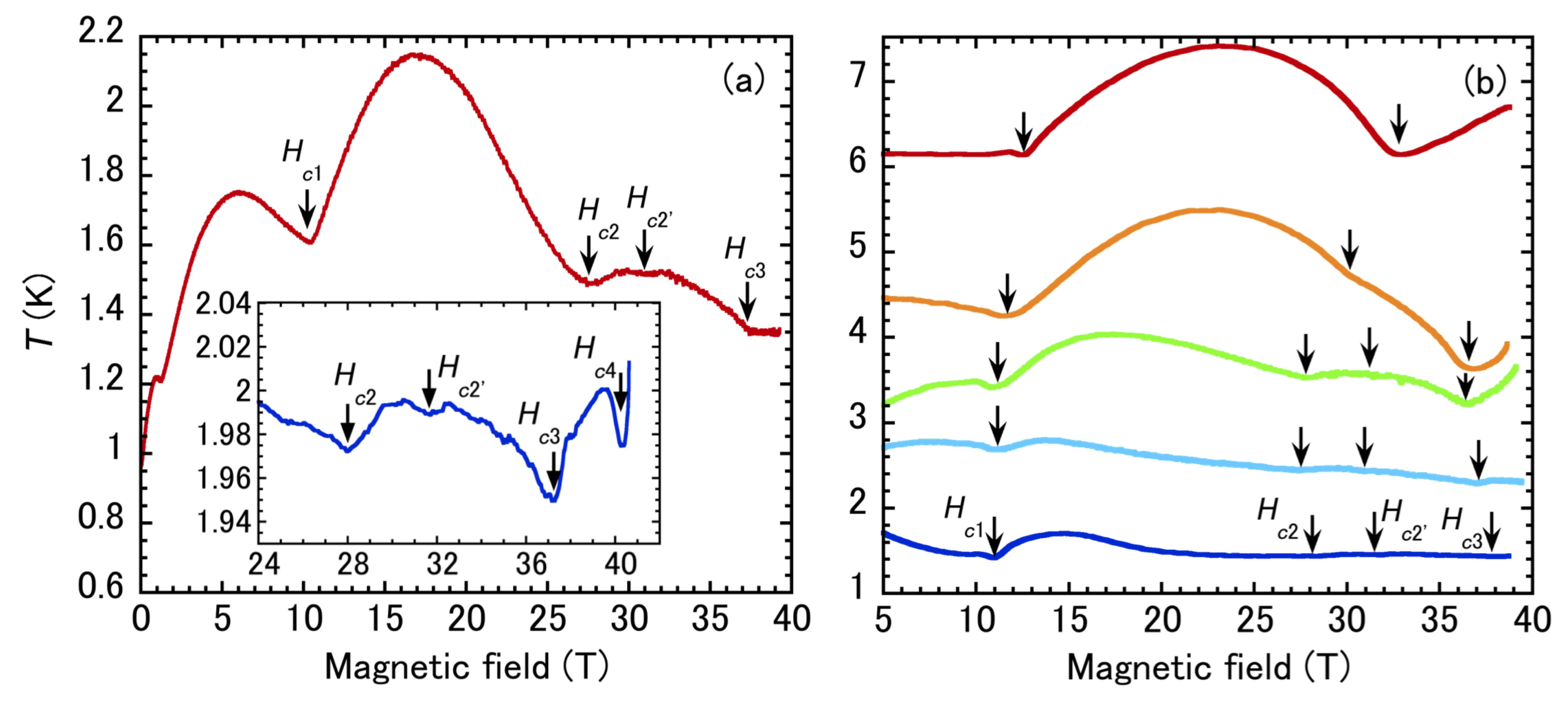}
\caption{Magnetocaloric effect of HgCr$_2$O$_4$, observed under (a) quasi-adiabatic and (b) quasi-isothermal conditions. The inset shows  the result of the magnetocaloric effect measurement under a condition intermediate between quasi-adiabatic and quasi-isothermal.}
\label{Frequency-field plot}
\end{figure}

\begin{figure}
\includegraphics[width=15cm,clip]{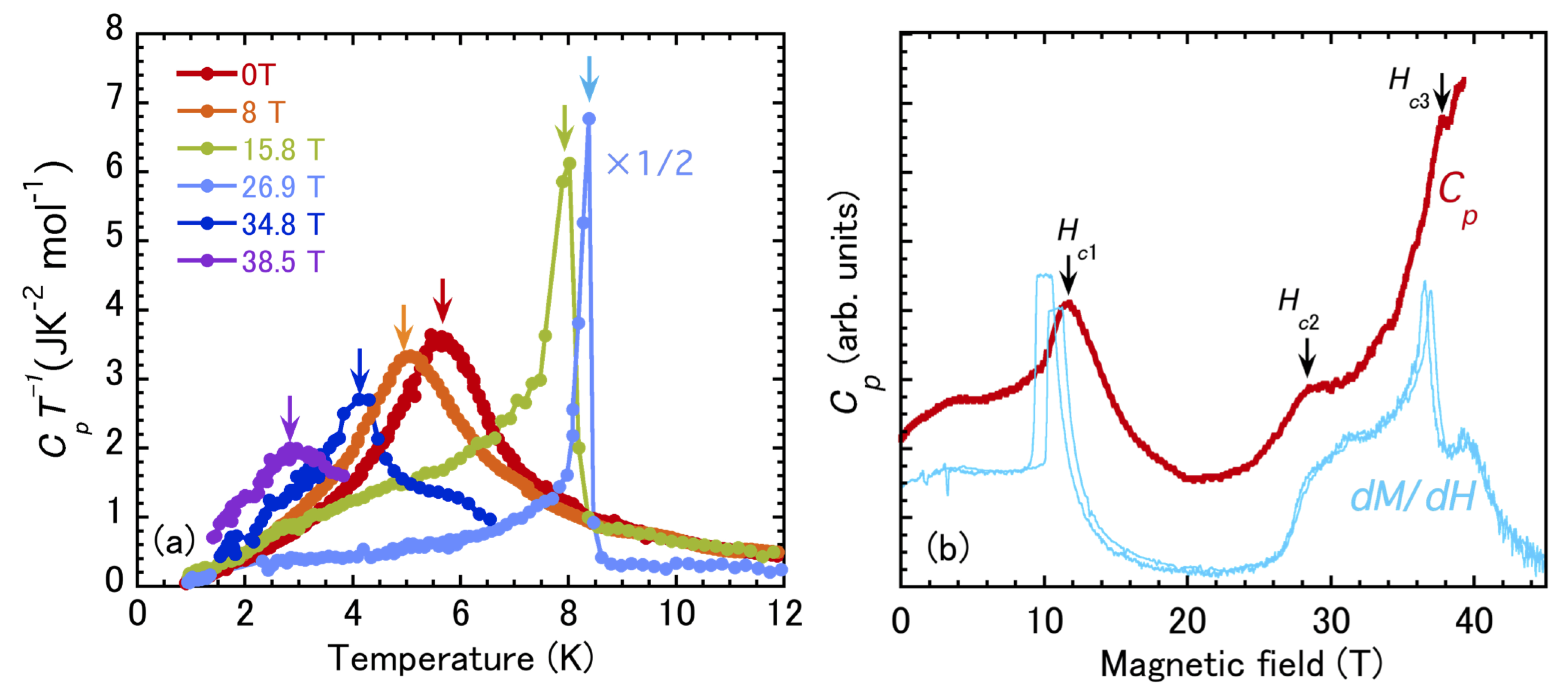}
\caption{(a) Temperature dependence of the specific heat $C_p$ decided by temperature $T$. The data at 26.9 T are multiplied by 1/2. (b) Magnetic field dependence of $C_p$ of HgCr$_2$O$_4$, observed at 1.3 K. Thin solid curves are the differential magnetization ( $dM$/$dH$) curve. }
\label{Frequency-field plot}
\end{figure}

\begin{figure}
\includegraphics[width=12cm,clip]{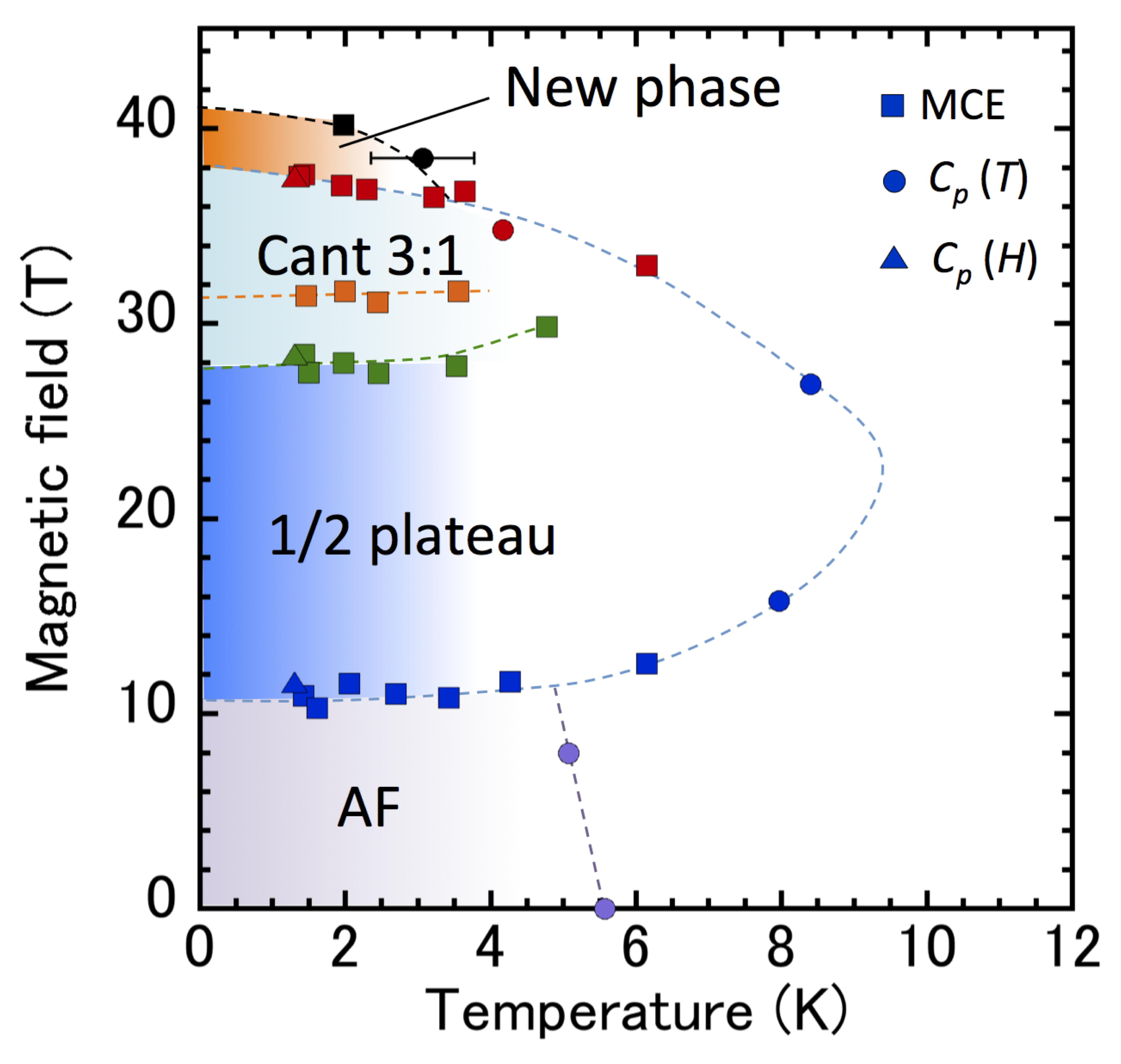}
\caption{Temperature to magnetic field phase diagram of HgCr$_2$O$_4$, obtained from thermodynamic experiments. Squares, circles, and triangles are the transition points obtained from the magnetocaloric effect, temperature, and field dependenceis of the specific heat measurements, respectively. Dashed curves are drawn as guides for the eyes.}
\label{Frequency-field plot}
\end{figure}

\end{document}